 \documentclass{ws-procs}
\begin{document}


\font\zfont = cmss10 
\font\litfont = cmr6
\font\fvfont=cmr5
\def\bigone{\hbox{1\kern -.23em {\rm l}}}  
\def\ZZ{\hbox{\zfont Z\kern-.4emZ}}
\def\hf{{\litfont {1 \over 2}}}
\def\mx#1{m_{\hbox{\fvfont #1}}}
\def\gx#1{g_{\hbox{\fvfont #1}}}
\def\Re{{\rm Re ~}}
\def\Im{{\rm Im ~}}
\def\lfm#1{\medskip\noindent\item{#1}}
\def\p{\partial}
\def\a{\alpha}
\def\b{\beta}
\def\g{\gamma}
\def\d{\delta}
\def\e{\epsilon}
\def\th{\theta}
\def\vt{\vartheta}
\def\k{\kappa}
\def\l{\lambda}
\def\m{\mu}
\def\n{\nu}
\def\x{\xi}
\def\r{\rho}
\def\vr{\varrho}
\def\s{\sigma}
\def\t{\tau}
\def\z{\zeta }
\def\vp{\varphi}
\def\G{\Gamma}
\def\D{\Delta}
\def\T{\Theta}
\def\X{\Xi}
\def\P{\Pi}
\def\S{\Sigma}
\def\L{\Lambda}
\def\O{\Omega}
\def\oo{\hat \omega }
\def\ov{\over}
\def\o{\omega }
\def\bbox{{\sqcap \ \ \ \ \sqcup}}
\def\tria{$\triangleright $}
\def\dlr{\buildrel \leftrightarrow \over \partial }
\def\ddlr{\buildrel \leftrightarrow \over d }
\def\Gsl{{G }}
\def\Csl{{C }}
\def\partialsl{\partial \ \ \ /}
\def\sn{{\rm sn} }
\def\L{\Lambda}

\def\QG{2D QG}
\def\t{\tau}
\def\g{\gamma }
 \def\CM {{\mathcal{M}}} 
  \def\CA {{\mathcal{A}}} 
  \def\CV {{\mathcal{V}}} 
  \def\CB {{\mathcal{B}}} 
  \def\CD {{\mathcal{D}}} 
  \def\CL {{\mathcal{L}}} 
   \def\CZ {{\mathcal{Z}}} 
   \def\R{\mathbb{R}}
 \def\p{\partial } 
 \def\d{\delta }
 \def\G{\Gamma }
 \def\<{\langle\,}
\def\>{\,\rangle}
 \def\({\left( }
\def\){\right) }
 \def\[{\left[ }
\def\]{\right] }
\def\hf{{\textstyle{1\over 2}}}
\def\s{\sigma}
\renewcommand{\theequation}{\thesection.\arabic{equation}}
\newcommand{\eqal}[2]{\begin{eqnarray} #2 \end{eqnarray}}
\newcommand{\eqarr}[2]{\begin{array} #2 \end{array}}
\newcommand{\eqn}[2]{\begin{equation} #2 \end{equation}}
\newcommand{\newsec}[1]{\setcounter{equation}{0} \section{#1}}
\newcommand{\no}{\nonumber}
\newcommand{\tr}{{\rm tr~}}
\newcommand{\Tr}{{\rm Tr~}}

\title{Boundary Ground Ring  and  Disc Correlation  Functions
  in Liouville  Quantum Gravity\footnote{Contribution
 to the proceedings of the conference
{\em Lie theory and its applications in physics - 5},
June 2003, Varna, Bulgaria } }

\author{Ivan Kostov}
\address{ Service de Physique 
Th{\'e}orique,
 CNRS -- URA 2306, C.E.A. - Saclay,   
  F-91191 Gif-Sur-Yvette, France;\\
  Associate member of  INRNE - BAS,  
Sofia, Bulgaria\\
{\tt kostov@spht.saclay.cea.fr}
}

%

%




\maketitle

\abstracts{
We  construct    the  boundary ground ring  in
$c\le 1$  open string theories
 with non-zero boundary cosmological constant
 (FZZT brane), using the Coulomb gas  representation.
 The ring  relations  yield an  over-determined set of functional  
recurrence equations for the boundary correlation functions, which involve shifts of the the target space 
momenta of the 
boundary fields as well as  the boundary parameters on the different 
segments  of the boundary. 
}


\section{Introduction}
  
  \def\QG{2D QG}
  
\noindent
The non-critical string theory, known also as 2D quantum gravity, has
two complementary descriptions: the world sheet description in terms
of Liouville CFT and the target space description in terms of random
matrix models (for reviews see \cite{GM, DiFrancescoGinsparg,
polchinski, KlebanovMQM, JevickiQN}).   It has been observed in the
early 90's that both descriptions possess integrable structures, the
KP and Toda integrable hierarchies for the matrix models and the
`ground ring' for the world sheet CFT. There are some similarities
between the integrable structures on both sides, which might be used,
if well understood, to build a more direct connection between the two
dual descriptions of 2D string theory more direct.  Such a connection
could be very helpful when studying strong perturbations of the 2D
string theory as the 2D Euclidean black hole.

 The integrable structure of the world sheet CFT \cite{ Witten, WitZw,
 KlebPol} is based on the so called ground ring of operators.  It has
 been shown by Witten \cite{Witten} that there exists a ring of
 dimensionless ghost-number zero operators, with respect to which the
 local observables form a module.  The action of ground ring leads to
 to a set of recurrence equations \cite{KMS} for the correlation
 functions of the primary fields (the closed string `tachyons').  It
 has been also observed \cite{Witten, KMS} that the ground ring
 relations resemble the string equations in the corresponding matrix
 models\footnote{This analogy explained and developed further in the
 recent paper \cite{SeibergS}.  }.  The ground ring structure was
 generalized in \cite{bershkut} to open string theory, described by a
 boundary CFT. The action of the `boundary ground ring' yields a set
 of recurrence relations for the correlation functions of boundary
 operators, or open string `tachyons'.  These (over determined)
 relations reproduced elegantly the results obtained previously by
 Coulomb gas integrals \cite{Gouli, VDotsenko, berkut, TY}.

 Originally  the   ring relations    were   applied  only for the 
 so called   `resonant'
amplitudes. These are the correlation functions for 
vanishing   cosmological constant, when the Liouville interaction 
is  not taken into account.
In the full theory, the resonant amplitudes give the residues of the 
`on-mass-shell' poles of the exact amplitudes. 
 It was    conjectured   \cite{Witten}  that  the  perturbations like 
the 
cosmological term 
   are described by  certain  deformations of the  ground ring 
structure,  
   but    the exact field-theoretical meaning of this conjecture 
   remained obscure.
Only after the   impressive developments in Liouville CFT  from
     the last several years  \cite{DO, ZZtp, FZZb, PTtwo, hosomichi}
     it  has been realized  that the  ground ring structure  is 
   an {\it exact}  symmetry of the full CFT \cite{newhat, SeibergS} .
      The  recurrence relations  that follow from the ground ring 
      are the string theory version of the relations obtained by the 
      conformal bootstrap in Liouville theory.

     In these notes we will focus on   boundary    2D quantum gravity,
  which  
    is characterized by two cosmological 
constants: 
      the bulk constant $\mu$ coupled to the area of the world sheet 
and the 
      boundary constant $\mu_B$ coupled to the length of the boundary.
      In the CFT description these  are the coupling  constants for 
the bulk and boundary   Liouville 
      interaction.
         In Liouville theory the boundary term can be   as well
introduced  
         as a non-homogeneous  Neumann boundary condition 
\cite{FZZb}, which is also referred to as FZZT brane.

  A   deformation of the  
{ boundary ground ring}  structure by  bulk and boundary Liouville 
interactions 
 was  proposed in \cite{BGR}. 
It was shown that the ring structure yields  a set of  {  functional }
  recurrence equations, which  completely determine 
 all boundary correlation functions on a  disc.
  Crucial role in  the derivation  plays   the precise definition of
  Liouville  degenerate boundary primaries  given  in  
\cite{FZZb}, which we  sketch  below.

The 
   observables in Liouville CFT with boundary 
      are meromorphic functions 
of  the boundary cosmological constant $\mu_B$  
    with  a cut along the semi-infinite
 interval $[-\infty, -\mu_B^0]$, where $\mu_B^0\sim\sqrt{\mu}$. 
  The branch point singularity can be resolved by 
 the uniformization  map
$$
 \mu_B = \mu_B^0 \cosh (\pi bs),
 $$
 where $b$ is the Liouville coupling constant
 and $s$ is a dimensionless parameter.
 It  happens that it is more appropriate to 
    label the Liouville boundary conditions 
  by the values of $s$  instead of $\mu_B$. 
Thus a  Liouville boundary 
operator  is   specified  by three numbers: the Liouville charge $\beta$
and the two boundary parameters  $s_{\rm left}$ and  $s_{\rm right}$.
 In \cite{FZZb}, Fateev, Zamolodchikov and Zamolodchikov (FZZ)  observed  
 that the correctly defined lowest degenerate boundary operators 
 introduce shifts of the  boundary parameter $s\to s\pm ib$ or $s\to 
s\pm  i b^{-1}$
  at the points of the boundary where  they are inserted\footnote{
The general validity of this suggestion, which was  proven in 
\cite{FZZb}\ 
only for the two-point function,  follows   from the  formula for the 
three point                                                          
function derived  subsequently in \cite{PTtwo}.}.
  The fusion  rules  with such an operator yield  a
  functional  equation   for the  boundary 
two-point function (see the 
 concluding section of \cite{FZZb}),   
 which can be written as a {\it difference equation}
with respect to one of the boundary parameters. 
  It was subsequently found    \cite{KPS}  that 
all boundary Liouville structure constants satisfy  
 similar  difference equations.

 In these notes we  review and slightly generalize the 
 result of  \cite{BGR} (there the  matter field obeys $U(1)$ fusion rules)
 by adding  the two screening charges to the effective action.
    The ring relations lead to a 
   set of functional  recurrence identities for all  boundary correlators, 
   from which we derive a  pair of difference equations with respect to one of the 
   Liouville boundary parameters.
   We will see that the  screening charges  add an extra term to the
    functional equations  of  \cite{BGR}, but the difference equations 
    remain the same.

 \def\LDE{linear difference equations }

\section{Preliminaries: world-sheet CFT  for the disc amplitudes }

  \def\nienhuis{\cite{nienhuis}}
  
 
 \subsection{The effective action}

\noindent
 We  consider  the  gaussian field realization  \cite{DF} of the matter CFT,
 which is very convenient for actual calculations.
  Then the Euclidean 2D string  theory   is defined in terms of 
    a Liouville  field $\phi$    with background charge $Q$ and  
    a gaussian  matter field $\chi$
   with background charge $e_0$.   
  The  simplest form of the  effective  action is achieved by
  mapping the disc   to the upper half
 plane $\{ \Im x\ge 0\}$. 
 Then the bulk and boundary curvatures are concentrated at infinity
 and  the effective action takes the   form
 \cite{FZZb,Teschner}
  \eqn\actg{ \CA[\phi, \chi] =\int \limits_{\Im x\ge 0}d^2 x\(
{1\over 4\pi} [ (\nabla \phi)^2 + (\nabla \chi)^2 ] +\mu e^{ 2b\phi 
}\)
+\int\limits_{-\infty}^\infty dx \ \mu_B\ e^{b\phi} + {\rm ghosts}
  \label{actg} }
 The couplings    $\mu $ and $\mu_B$ are  are  
 referred to as bulk and the boundary
cosmological constants.
The  matter and Liouville background charges 
 \eqn\eoq{
 e_0= {1/b} - b,   \quad Q= {1/b}+b}
are introduced through the asymptotics of
the fields at   infinity \eqn\asyms{
\phi(x, \bar x ) \sim - Q \log |x|^2
,\quad \chi (x, \bar x) \sim - e_0\log |x|^2.
}
The two background charges satisfy $Q^2 -
 e_0^2=4$, which is equivalent to the balance of the central charge $
 c_{\rm tot}\equiv c_\phi +c_\chi+ c_{\rm ghosts}=(1+6 Q^2)+(1-
 6e_0^2) -26 =0.  $ 
 
The  boundary term encodes a inhomogeneous 
Neumann boundary condition for the Liouville field
\eqn\bcphi{
  i(\p-\bar \p) \phi(x, \bar x)  =4\pi \mu_B \ e^{b\phi(x, 
\bar x )}\qquad
(\bar x =x),}
the so called  FZZT brane.
  The matter field 
   is assumed to satisfy pure Neumann
 boundary condition:
 \eqn\bcchi{ i(\p-\bar \p) \chi(x, \bar x)  =0 \qquad
(\bar x =x),}
  which is of course  Dirichlet boundary condition for the 
 T-dual field $\tilde \chi (x, \bar x)
  .$

  
The Coulomb gas mapping requires also to add to the action 
(\ref{actg})  the  
 bulk and boundary screening charges
\eqal\scrch{
 Q_+ = \mu _+ \int _{\Im x\ge 0} d^2 x\  e^{-2ib\chi} , \ \ \ \
 Q_- = \mu_-  \int _{\Im x\ge 0} d^2 x   e^{2i\chi/b} ) ,
 \label{scrch}}
\eqal\scrchb{
Q_+^B =   \mu ^B_+  \int _{-\infty } ^{\infty} dx 
e^{-ib\chi} , \ \ \ \
 Q_- ^B=  \mu^B_- \int _{-\infty } ^{\infty} dx  \
e^{i\chi/b} 
\label{scrchb}}
The  values of the  `couplings'  $\mu_\pm$ and $\mu_\pm^B$ 
 are related to  the normalization of the correlation functions.
 Below we
will   choose particular values for which  the fusion rules 
acquire  simplest form.

\subsection{Bulk and boundary vertex operators }

 \noindent
  The bulk primary fields (closed string tachyons)
$\CV_P^{(\pm)} (x,\bar x)$ are defined as
  the bulk vertex operators
\eqal\newnor{ 
 \CV_P^{(+)}&=  \frac{1}{\pi}
 \gamma( bP)\ 
  e^{i(e_0-P)\chi  +(Q- P)\phi}\cr 
 \CV_P^{(-)}&= {  1\over \pi}
 \gamma\(- \frac{1 }{b}P\) \ 
  e^{i(e_0-P)\chi  +(Q+ P)\phi}
\label{newnor}
 }
 where $\g(x)\equiv  \G(x)/\G(1-x)$. 
 This normalization removes the external  `leg pole' factors   
in the correlation functions and  is the one to be used  when  
comparing with the microscopic realizations of \QG.   
%
The boundary primary fields $\CB _P ^{(\pm)} (x)$,  or 
   left/right moving  on-mass-shell open   string tachyons, 
   are defined as the 
boundary vertex operators
 \eqal\normB{ 
 \CB  _P^{(+)}
   &={1\over \pi} \  \G(2b P)\ 
   e^{ i \(\hf e_0-P\) \chi+\(\hf Q - P\) \phi}\cr
\CB  _P^{(-)}
  & ={1\over \pi}  \ \G\(-\frac{2}{b}P\)\ 
   e^{ i \(\hf e_0-P\) \chi+\(\hf Q + P\) \phi}.
    \label{normB}
    }
As any    CFT boundary operator,  the open string tachyon is  
completely 
 defined only after  the  left and right boundary conditions are 
specified \cite{cardy}.   
The matter component   is assumed to satisfy  homogeneous Neumann 
boundary condition, and 
 the Liouville left and right boundary conditions 
are determined by 
the 
values  of the 
uniformization parameter $s$ 
 defined
 as
  \eqn\deftau{
 \mu_B = \mu_B^0 \cosh (\pi bs).
 \label{deftau}
 } 
  Following  \cite{FZZb},  we will denote  such an operator
  as 
 $^{s_{1}} [\CB  _P^{(\pm)}]^{s_{2}}$. 
    
    The tachyons of opposite chiralities are related by the Liouville  
bulk and boundary reflection amplitudes 
\eqal\reflBul{ 
\CV_P^{(+ )}= S_P^{(+-)} \ \CV_P^{(-)}, \qquad 
^{s _1}  [\CB^{(+)}_P]^{s _2} =  D_P^{(+-)}  (s _1, s _2) 
\ ^{s _1}  [\CB^{(-)}_P]^{s _2}.
\label{reflBul}
} 
  For each value of the momentum there is only one 
  vertex operator that corresponds to a physical state. 
  The physical operators are
  \eqal\physo{
   \CV_P = 
   \begin{cases}
    \CV  _P^{(+)}      & \text{if  } P<0, \\
       \CV  _{P}^{(-)} & \text{if }  P>0 .
\end{cases}
\qquad      
  \CB_P= \begin{cases}   \CB  _P^{(+)}     & \text{if  } P<0, \\
   \CB  _{P}^{(-)}\
  & \text{if }  P>0 .
\end{cases}
 \label{physo} }
  The  ``wrongly dressed" operators are related to the physical ones 
by  (\ref{reflBul}).

\subsection{Normalization of the bulk and boundary cosmological 
constants and the screening operators}

\noindent
It is  convenient to redefine the   
cosmological constants    $\mu\to \L$ and $\mu_B\to z$   according to
the normalizations   \ref{newnor}    of the vertex 
operators :
  \eqn\intBB{
\mu    e^{2b\phi} =
\Lambda\  \CV^{(+)}_{e_0},
\quad \mu_B \ e^{b\phi}= z\  \CB  ^{(+)}_{e_0/2}.  
\label{intBB}
  }
 The new cosmological constants are related to the old ones by
 \eqn\Mmu{
{\Lambda}  =\pi \g(b^2) \mu ,
\qquad {z} =  {\pi [\G(1-b^2)]^{-1}} \mu_B.
  \label{Mmu}
} 
Now the  uniformization map \ref{deftau}\  reads%
\eqn\zofM{
 z = M\cosh (\pi b s)  , \qquad M= \sqrt{\Lambda}.
 }
Also, it will be  convenient to normalize the screening charges as
\eqal\nscrc{
Q_\pm = \int \limits_{\Im x>0} d^2 x\ \CV^{(\pm) } _{\pm  Q}, \quad
Q_\pm^B = \int \limits_{-\infty} ^{\infty} dx \ \CV^{(\pm) }_ {\pm Q/2}\
\label{nasrc}.}

 In the normalization (\ref{intBB}) 
 the self-duality  property  \cite{FZZb}
 of Liouville quantum gravity reads
 $$ b\to \tilde b = 1/b, \qquad \L\to \tilde\L,  \qquad  M\to \tilde M,
 $$
  with
   \eqn\DUALTY{ \tilde \Lambda =
 \Lambda^{1/b^2} \qquad {\rm or}\qquad \tilde M= M^{1/b^2}.
} 
The boundary parameter $s$ is self-dual.  The duality transformation
for the correlation functions follows from the duality transformation
of left and right chiral tachyons
   \eqn\TDU{ \tilde \CV ^{(\pm)}_{P} =
\overline{ \CV ^{(\mp)}_{-P}} , \qquad ^{s_1}[\tilde \CB ^{(\pm)}_{P}
]^{s_2} =\ ^{s_1}\![ \overline{ \CB ^{(\mp)}_{-P}}]^{s_2}, 
   \label{TDU}}
where the bar means   complex conjugation.

  \subsection{Degenerate fields}

  \noindent
   By degenerate fields in Liouville quantum gravity we will 
understand 
   the gravitationally dressed degenerate matter fields.
    In our case these are the on-shell vertex operators  (\ref{physo})
    with  degenerate matter components
\eqal\degenPrs{ 
 \CV_{rs}=    \CV_{r/b-sb}  , \qquad \CB_{rs} = \CB_{(r/b-sb)/2}
 \qquad (r,s\in N).
 }
 The flat scaling dimensions of these fields are
 $$ \Delta_{rs}= {(r/b-sb)^2 - e_0^2\over 4}.$$
   The fusion rules for the degenerate  in Liouville quantum gravity 
    are  the same as
 the fusion rules for the degenerate fields  in the matter CFT \cite{BPZ}.
 Note that  the  Liouville components of these fields   are not
   degenerate  Liouville fields.
 
    
\section{Bulk ground ring}

\subsection{The ring relations}


\noindent
Before considering the boundary ground ring, we 
   will give  a  short review of  the   ground ring structure
for a theory without boundary.  
     The    operators  that span the ground ring are obtained by
     applying a Virasoro raising  operator
of level $rs-1$ to the
        product of  two degenerate matter and 
Liouville fields
with Kac labels $r,s$.
   The resulting operators have conformal 
weights $\Delta=\bar\Delta=0$.
The    ring is generated by the first  two lowest operators 
\cite{Witten}
\eqal\bulkGR{
a_+=&-|{\rm\bf bc}- b\p_x(\phi-i \chi)|^2\
e^{- b^{-1}(\phi+i\chi)}\cr
 a_-=&-  |{\rm\bf bc}- b^{-1} \p_z(\phi+i \chi)|^2\
 e^{ -b(\phi -i\chi)}
 }
where  ${\rm\bf  b, c}$  are the
reparametrization ghost and anti-ghost fields.
The  ground ring is spanned on the polynomials
 $(a_+)^m(a_-)^n$ with  $m,n\in Z_+$.
 In the case of non-rational $b^2$ the ground ring contains no other 
relations
 and has an infinite number of elements labeled by the integers 
$r,s\ge 1$.    

 A crucial property of the operators $a_\pm$ is that their
 derivatives $\p_x a_\pm$ and  $\p_{\bar x} a^\pm$ are
    BRST exact: $\p_x a_- = \{Q_{\rm BRST}, {\rm\bf b}_{-1} a_-\}$. 
    Therefore, any amplitude that
involves $ a_\pm$  and other BRST invariant 
operators does not depend on the
position of $ a_\pm$.
This property allows to  write recurrence equations for the 
correlation functions from the  OPE   of $a_\pm$ and the other 
BRST-invariant fields.

The vertex operators   (\ref{newnor})  form a module under the
  ground ring:
\eqn\actaa{
 a_{+} \CV_P^{(+)} = -
   \CV_{P+  {1\over b } }^{(+)} , \qquad
   a_{-}  \CV_P^{(-)} =- 
   \CV_{P-  {b} }^{(-)} .
\label{actaa}}
  and also 
  \eqn\actaab{
  a_{+} \CV_P^{(-)} = a_{-} \CV_P^{(+)} =0.
  \label{actaab}}
  Both relations  (\ref{actaa}) and  (\ref{actaab})  follow from the 
free 
field OPE and 
  are true up to commutators  with the BRST charge.
 While the first one survives in the  correlation functions,
 the second one receives non-linear corrections. 
 Due to the contact terms, the last relation  should be 
  modified   in presence of an integrated  vertex  operator 
  to  
  \eqn\aplus{
    a_+  \CV^{(-)}_P\int d^2 z\   \CV^{(-)}_{P_1}
=  
\CV^{(-)}_{P+P_1+ b}
\label{aplus}}
\eqn\aminus{
 a_-  \CV^{(+)}_P\int d^2 z \   \CV^{(+)}_{P_1}
=   
\CV^{(+)}_{P+P_1-1/b.}
\label{aminus}}
These relations  imply
 a set of recurrence equations  \cite{KMS, bershkut, kachru}, which
   determine completely the resonant   
tachyon amplitudes. 
 
After taking into account the  Liouville interaction and the screening charges 
second  relation   (\ref{actaab})  gets  deformed.
 Remarkably, the  deformation    can be calculated pertubatively
  by  applying (\ref{aplus}) and (\ref{aminus}):
      \eqal\actaD{
a_-  \CV^{(+)}_P 
&=&   - \L\ 
\CV^{(+)}_{P-  b } +\CV^{(+)}_{P+  b } 
\cr
a_+  \CV^{(-)}_P &=&    - \tilde \L\ 
\CV^{(-)}_{P+1/  b } +  \CV^{(-)}_{P-1/  b }.
\label{actaD}}
 It is not at all evident that the linear insertions 
 allow to calculate  the exact action 
 of  $a_\pm$, but this  has been justified by various  self-consistency checks 
in  Liouville theory \cite{DO, ZZtp}.
 The  ground ring structure  in presence of screening charges has 
been considered
 in    \cite{
 GovJcf, 
 GovLast}.

\subsection{Functional equations for the bulk corrrelators}

Using  the ring relations  (\ref{actaa}) and  (\ref{actaD})
 one   can obtain a set of recurrence  equations 
 for the   correlation functions of the bulk tachyons
 \eqal\defG{&
 G(P_1,...,P_n|P_{n+1},...,P_{n+m})=
 \Big < \[\prod_{k=1}^{n-1}
 \int 
  \CV^{(-)}_{P_k}\] {\rm\bf c\bar c}  \CV^{(-)}_{P_n}(0) \times 
  \cr &\times 
{\rm\bf c\bar c}\CV^{(+)}_{P_{n+1}}(1)  
\[
  \prod_{j=n+2}^{n+m-1}
  \int 
\CV^{(+)}_{P_{j}}
 \]
  {\rm\bf c\bar c} \CV^{(+)}_{P_{n+m}}(\infty)
\Big>.
 }
 Consider the auxiliary function 
 \eqal\auxF{
 &F(x,\bar x |P_1,...,P_n|P_{n+1},...,P_{n+m})=
 \< \[ \prod_{k=1}^{n-1}\int
  \CV^{(-)}_{P_k} 
  \]
   {\rm\bf c\bar c} \CV^{(-)}_{P_n+b}
   (0)\times \cr & \times  a_-(x,\bar x)\  {\rm\bf c\bar c} \CV^{(+)}_{P_{n+1}}(1)
  \[\prod_{j=n+2}^{m+n-1}\int {\rm\bf c\bar c}\CV^{(+)}_{P_{j}}
 \]  {\rm\bf c\bar c} \CV^{(+)}_{P_{n+m}}(\infty)  \>
 \label{auxF}}
 The function $F$ 
does not depend on $x$ and $\bar x$.  This can be proved by   using
$\p_x a_-=\{Q_{\rm BRST}, {\rm \bf b}_{-1} a_-\}$
and deforming the contour, commuting $Q_{\rm BRST}$ with the other 
operators
in (\ref{auxF}) \cite{bershkut}. Therefore one can
evaluate this function at $x=0$ and $x=1$ by
using the fusion rules  (\ref{actaa})   and  (\ref{actaD}). 
As a result one obtains the recurrence relation
      \eqal\recCL{
      &G(P_1,...,P_n|P_{n+1},...,P_{n+m})    =   \L \   
     G(P_1,...,P_n+b|P_{n+1}-b,...,P_{n+m})-\cr &
+
     G(P_1,...,P_n-b|P_{n+1}-b,...,P_{n+m})\cr &
     -\sum_{j=1}^{m-1} 
     G(P_1,...,P_n+b |P_{n+2},
     ... , P_{n+j} +P_{n+m} -\frac{1}{b},  ...,P_{n+m})
     .}  
 Similarly, by inserting   $a_+$    
 we get  the dual recurrence relation
    \eqal\recCLd{
      & G(P_1,...,P_n|P_{n+1},...,P_{n+m})    = 
       \tilde  \L \   
     G(P_1,...,P_n+\frac{1}{b}|P_{n+1}-\frac{1}{b} ,...,P_{n+m})\cr 
   &+ \  G(P_1,...,P_n-\frac{1}{b}|P_{n+1}-\frac{1}{b} ,...,P_{n+m})
   \cr&   -
      \sum_{k=1}^{n-1} 
      G(P_1,...,P_k +P_n+b, ..., P_{n-1}|P_{n+1},...,P_{n+m}).}     
          Note that the three-point function  $m+n=3$ coincides with the 
  corresponding  Liouville  three-point function.
  The latter has been evaluated     by using a 
very similar argument   in   \cite{Teschner}.

\newsec{The boundary   ground ring}

\def\inv{^{-1}}

\subsection{The  ring relations}

\noindent
Similarly to the bulk case the  two generators of the boundary 
ground ring are defined as
 \eqal\APAM{
A_+
& 
=- [{\rm bc} - \hf b\p_x(\phi -i\chi) ] \ e^{-\hf b\inv  
(\phi+i\chi)}\cr
A_- 
&  
=- [{\rm bc} - \hf b^{-1} \p_x(\phi +i\chi) ] \ e^{-\hf b 
(\phi-i\chi)}.
\label{APAM}}
where ${\rm b, c}$ are the boundary  reparametrization ghosts.
The two operators are related by a duality transformation combined 
with complex conjugation:
\eqn\conjA{
\tilde A_+ = \overline{A}_-, \qquad
\tilde A_- = \overline{A}_+.
\label{conjA}}
These operators  are BRST closed: $ \p_x A_\pm = \{ Q_{\rm BRST} , {\rm b}_{-1} 
A_\pm\}$ 
and have $\Delta=0$. The  open string tachyons (\ref{normB})
form a module with respect to the 
ring generated by these two operators.

 Let us   first  consider the  simplest case of
 pure Neumann b.c.   for  the  Liouville  field ($\mu_B=0$),
 following 
\cite{bershkut}. 
Then the 
action of the   ring on the tachyon modules is 
generated by the relations
 \eqn\ACTAA{
 A_{+} \CB  _P^{(+)} = 
  \CB  _{P+  {1\over 2 b } }^{(+)} ,\qquad 
   A_{-} \CB  _P^{(-)} = 
  \CB  _{P-  {b\over 2} }^{(-)} 
 \label{ACTAA} }
    \eqn\ACTAAB{
  A_{+} \CB_P^{(-)} = A_{-} \CB_P^{(+)} =0  .
  \label{ACTAAB}}
   Again, the first relation (\ref{ACTAA})  is exact and the second 
relation (\ref{ACTAAB})
  gets deformed in presence 
   of integrated boundary tachyon fields  \cite{bershkut} :
   \eqn\nonAr{ A_- \CB  _{P_1}^{(+)} \int dx \CB  _{P_2}^{(+)}=
   {1\over \sin 2\pi b P_1 }\   \CB  _{P_1+P_2-  {1\over 2b}}^{(+)}.
      \label{nonAr}}
     \eqn\nonAl{
    \int dx \CB  _{P_1}^{(+)} A_- \CB  _{P_2}^{(+)} =
    {\sin 2\pi b (P_1 +P_2)\over \sin 2\pi b P_1\ 
    \sin 2\pi bP_2}\ 
     \CB  _{P_2+P_1-  {1\over 2b}}^{(+)}.
  \label{nonAl} }
   \eqn\nonApr{A_+
   \CB  _{P_1}^{(-)} \int dx\CB  _{P_2}^{(-)}=
   {1\over \sin {2\pi\over b}   P_1 }\  
    \CB  _{P_1+P_2+  {b\over 2}}^{(-)}
  \label{nonApr} }
    \eqn\nonApl{
    \int dx \CB  _{P_1}^{(-)} A_+\CB  _{P_2}^{(-)} =
    {\sin {2\pi\over b}  (P_1 +P_2 )\over 
    \sin {2\pi\over b} P_2 \ \sin {2\pi\over b}  P_1 }\ 
     \CB  _{P_1+P_2+ {b\over 2}}^{(-)}.
      \label{nonApl}   }
       The coeffficients on the r.h.s. are obtained as standard Coulomb integrals.   As in the bulk case,  these relations  yield 
an over-determined set of recurrence equations for 
        the  $n$-point  open string amplitudes \cite{bershkut}.

 \subsection{Deformation of the ring relations 
 by the boundary Liouville  
 term and the screening charges}

    With   the  screening charges
   (\ref{nasrc})  added to   the effective action, 
   the     action of the ring on the vertex operators 
    gets deformed  according to the  general relations  (\ref{ACTAA}) and
          (\ref{nonAr}). Namely the relations
        (\ref{ACTAAB})  become
        \eqal\ACTscr{ 
      A_- \CB^{(+)}_P = n_P\cdot  \ \CB^{(+)}_{P+{b\over 2}}, \quad 
      A_+\CB^{(-)}_P = \tilde n_P\cdot  \ \CB^{(-)}_{P-{1\over 2b}}, 
    \label{ACTscr}}
where the coefficients $m_P$ and $\tilde m_P$ are given by
 \eqal\mPmP{n_P& =& \tan (\pi b P) - \tan (\pi b e_0/2)\cr
 \tilde n_P &=& \tan (\pi  P/b) + \tan( \pi  e_0 /2b).
 }

\def\Cpm{ m_{_P} ^{\pm}}
\def\Cpmd{  \tilde m_{_P} ^{\pm}}
 
 The evaluation of the deformation of the  the ring  relations   by the 
boundary Liouville interaction is more subtle.
 It is based on the 
  observation made in  \cite{FZZb}
and subsequently confirmed in \cite{PTtwo, KPS} 
that a 
  level-$n$ degenerate boundary Liouville field  
  $e^{-nb\phi}$  has vanishing null-vector and therefore a truncated 
OPE with the other primary fields if either $s_{\rm left} - s_{\rm 
right } = i b k$ or 
  $s_{\rm left} + s_{\rm right } = i b k$, with $k = -n/2, 
-n/2+1,..., n/2$.
  By the duality property  of Liouville theory, the degenerate 
boundary  fields 
  $e^{-n \phi/b}$ exhibit a similar property    
  with $b$ replaced by $1/b$.   No direct proof is supplied for 
  this statement, but it was shown to be consistent with the exact   
results obtained in boundary Liouville theory. 
 In particular,    the above property leads to a pair of
   functional equations for the Liouville  boundary reflection 
amplitude, whose
unique solution coincides with the result of the standard conformal 
bootstrap  \cite{FZZb}.

  According to \cite{FZZb}, 
 the  operators (\ref{APAM}) should be   defined  by
 \eqal\APMnew{
 A_+\quad &\to\quad  ^s[A_+]^{s \pm i /b}\cr
 A_-\quad &\to\quad  ^s[A_-]^{s \pm i b}
 }
and the relations  (\ref{ACTAA}),   (\ref{nonAr}),  
  (\ref{nonApr})   and   (\ref{nonApl})  
understood as
    \eqn\leftA{
^{s_1}[ \CB  _P^{(-)} ] ^{s }  [A_-]^{s \pm i  b} = -\ 
 ^{s _1} [\CB  _{P-  {b\over 2} }^{(-)}]^{s \pm ib} 
  \label{leftA}}
  etc.
  Taking into account the   three possible insertions of the boundary Liouville 
  interaction   boundary interaction  one finds  \cite{BGR} that 
  the relations 
(\ref{ACTscr}) deform further to 
         \eqn\funceq{
  ^{s_1}[  A_- ]^{s _{1}\pm i  b}
 [\CB  _{P } ^{(+)}]^{  s _2}
 =  \Cpm\cdot 
   \  ^{ s _{1} }[\CB  _{P -{b\over 2}} ^{(+)}]^{ s _2}
   + n_P\cdot \ ^{ s _{1} }[\CB  _{P +{b\over 2}} ^{(+)}]^{ s _2}
   \label{funceq}}
     \eqn\funceqd{
  ^{s_1}[  A_+]^{s _{1}\pm i /b }
 [\CB  _{P } ^{(-)}]^{  s _2}
 =\Cpmd \cdot \    ^{ s _{1} } [\CB  _{P +{1\over 2b}} ^{(+)}]^{ s _2}+
     \tilde n_P \cdot   \ ^{ s _{1} } [\CB  _{P -{1\over 2b}} ^{(+)}]^{ s _2} 
      \label{funceqd} }
 with the coefficients $\Cpm = \Cpm(s_1, s_2)$ 
     and $\Cpmd = \Cpmd(s_1, s_2)$ given by 
   \eqal\cccc{
\Cpm
    =- M \ {\cosh[ \pi b (s _1\pm 2  i  P) ]+ \cosh(\pi b s _2)
  \over \sin 2\pi b P}.
    \label{cccc}
    }
 \eqal\cct{
  \Cpmd=-M^{1/b^2} \ 
   {\cosh[\pi   \( {s _1} \pm 2i P\) /b ]+ \cosh(\pi {s _2/ b})
   \over \sin 2{\pi\over b}  P}  .
  \label{cct}}
 The derivation of   (\ref{funceq}) and (\ref{funceqd}) 
  essentially  repeats  the one presented in \cite{FZZb} for pure Liouville vertex operators and leads to the same fusion coefficients.
  
         \subsection{Functional and difference   
         equations for the correlation functions
  of boundary primary fields}

    Consider  a  boundary correlation function of the form
           \eqn\defWP{
 W_{P_1, P_2, ... }(s _1, s , s _2,s_3, ...) 
 =\< ^{s _1}[\CB _  {P_1} ]^{s } [ \CB  _{P_2}  ] 
 ^{ s_2 }[\CB  _{P_3}  ]^{s _3} ...\>
\label{defWP}
}
   The realization of the physical boundary fields 
depends on the sign of the momenta and 
is given by (\ref{physo}).
         Our notations 
      do  not distinguish between integrated 
 and non-integrated fields; it is assumed that 
       three of the integrations should be replaced by
      ghost insertions.
The amplitude (\ref{defWP}) is  
  analytic in the boundary 
parameters $s, s_1,...$  but not in the target space momenta, since 
    the fields with positive and negative momenta 
    are realized by different vertex operators. 
By the momentum conservation the correlation function 
is zero  unless 
    $\sum_k (\hf e_0 - P_k) = e_0  + m/b - n b $
    for some  positive integers $m$ and $n$.
   Equation 
   (\ref{TDU})   implies   the identity
   \eqn\revss{ W_{P_1, P_2, ... }
   =  \overline{W}_{-P_1, -P_2, ... }  
   = W_{-P_1, -P_2, ... }  .
 \label{revss}
     }

 Let us assume  that
 $P_1<0, \ P_2>0$. 
 Then the amplitude (\ref{defWP}) is
 realized as
 \eqn\defWPa{
 W_{P_1, P_2, P_3, ... }(s _1, s , s _2,s_3, ...) 
 =\< ^{s _1}[\CB _  {P_1}^{(-)}]^{s } [ \CB  _{P_2} ^{(+)}] 
 ^{ s_2 }[\CB  _{P_3} ^{(+)}]^{s _3} ...\>}
 By the   symmetry  (\ref{revss})
  the amplitude 
   can be 
 written also  as 
 \eqn\defWPd{
  \overline{W}_{P_1, P_2,P_3,  ... }(s _1, s , s _2,s_3, ...) 
 =\< ... ^{s _3}[\CB _  {-P_3}^{(-)}]^{s_2 } [ \CB  _{-P_2} ^{(-)}] 
 ^{ s}[\CB  _{-P_1} ^{(+)}]^{s _1} \> .
 \label{defWPd}}
  We will use  these two representations  to derive two independent 
functional identities 
 for  $W$.

   Consider the   auxiliary correlation function 
 $F$  with an operator $A_-$ inserted in the r.h.s. of  (\ref{defWP}):
 $$F =
 \< ^{s _1}[\CB  _{P_1+{b\over 2}}^{(-)}]^{s }[  A_- ]^{s \pm ib}
[  \CB  _{P_2}^{(+)} ]^{  s _2}[\CB  _{P_3} ^{(+)}]^{  s _3}
 ...\>.
 $$
 Let us assume further  that $P_1< -{b\over 2}$ and $P_2>{b\over 2}$.
    Then, evaluating $F$  by   (\ref{leftA})  and by   (\ref{funceq}) 
 and equating the results 
   we get the following functional  identities
 for the correlation functions of   three or more fields:
   \eqal\FUE{
&  &   W_{P_1
, P_2 ,P_3,  ... }(s _1, s \pm ib  , s _2,s_3, ...)   =\cr 
& +&  m_{_{P_2}}(s, s_2)\cdot  \ 
  W_{P_1+{b\over 2}, P_2-{b\over 2},P_3, ... }(s _1, s   , s _2,s_2, 
...)\cr
&+& \  n_{_{P_2} }\cdot \ 
  W_{P_1-{b\over 2}, P_2-{b\over 2},P_3, ... }(s _1, s   , s _2,s_2, 
...)\cr
& +& 
{\sin^{-1} (2\pi b P_2)}\cdot  \
W_{P_1+{b\over 2}, P_2+P_3 -{1\over 2b}, ... }(s _1, s   ,s_3, ...).
\label{FUE}}
The last,  `contact', term represents a correlation function with 
one  operator less. 
  A dual equation can be obtained in the same way,
assuming  that $P_1<-{1\over 2b}$ and $P_2>{1\over 2b}$,
 by evaluating the 
  complex conjugated function $\overline{F}$ using
   the representation (\ref{defWPd}) and the relations (\ref{conjA}).

  The symmetric part of     (\ref{FUE})  
   generalize  the recurrence 
equations 
of    \cite{bershkut} derived in the free-field ($\mu_B=0$) limit. 
  The antisymmetric  part  equation (\ref{FUE})  and  
 its dual  represents  a pair  of { homogeneous}  
 difference equations which 
do not have analog in the   $\mu_B=0$ limit.  
   We write them  in operator form:
 \eqal\FUim{ 
 \[  \sin ( b \p_s)-M \sinh (\pi bs  )\
  e^{{b\over 2}(\p_{P_1} - \p_{P_2})}\]
 W_{_{P_1, P_2,P_3,  ... }}(s _1, s , s _2,s_3, ...)  \ \ \ &&\cr
   =0\ \ \ \ \ \ \ \ & &
\label{FUim}}
for  $P_1< -{b\over 2}$ and $P_2>{b\over 2}$, and
 \eqal\FUimD{ 
 \[  \sin ( \frac{1}{b }  \p_s)-M^{1\over b^2} \sinh (\frac{\pi}{b } s  )\
  e^{{1\over 2b}(\p_{P_1} - \p_{P_2})}\]
 W_{_{P_1, P_2,P_3,  ... }}(s _1, s , s _2,s_3, ...)  \ \ \ &&\cr
 =0\ \ \ \ \ \ \ \ &  &
\label{FUimD}
 }
for $P_1<-{1\over 2b}$ and $P_2>{1\over 2b}$.
 
 Note that these equations  have the same form for  
  matter CFT with or without screening charges.
 Difference equations for all values of the momenta  can be 
obtained  by the following rule \cite{Ibliou, KPS}.
  If  the sign of one or  both  momenta  changes after the shift,
one should  apply the reflection property with the
amplitude $D^{(+-)}_P(s_1,s_2)$.
The details are explained in \cite{KPS, BGR}.

  \section{Discussion}
  
We have constructed the boundary ground ring 
in the Coulomb gas realization of the matter CFT. 
 Since  any matter CFT can be constructed from 
 the Coulomb gas,  the validity of the difference equations
 (\ref{FUim}) and (\ref{FUimD}) is universal.  Unlike underlying functional equations, the  form of the difference equations 
 is not altered by adding the screening charges.
     Remarkably, these equations 
  have their equivalent in the  microscopic realization  
  of \QG\  as loop gas on  random planar graphs \cite{Idis}.
  In this realization the degenerate boundary operators are represented 
  geometrically as sources of open lines with endpoints at the boindary.
     As a matter of fact, the difference equations (\ref{FUim}) and  (\ref{FUimD})
      have been  first derived  in the microscopic approach  \cite{ Ibliou,KPS}    
   by cutting open the sum over triangulated surfaces 
    along one of the lines and then using  certain factorization property
    of the  sum over surfaces.  Of course,  these equations are equivalent to 
    certain word identities   in the corresponding matrix model. 
     This confirmes the  deep 
   relation between the integrable structures in  the world sheet and 
   microscopic formulations  of \QG.

\section*{Acknowledgments}

I would like to express my gratitude to Valya Petkova for  many 
discussions and suggestions  and  to all  organizers of   Lie Groups - 5 
  for  inviting me to  this nice   workshop.
This research   is supported in part by the 
 European network  EUCLID, HPRN-CT-2002-00325.

\def\cqg#1#2#3{{ Class. Quantum Grav.} {\bf #1} (#2) #3}
\def\np#1#2#3{{Nucl. Phys.} { #1} (#2) #3}
\def\pl#1#2#3{{Phys. Lett. }{B#1} (#2) #3}
\def\prl#1#2#3{{Phys. Rev. Lett.}{\bf #1} (#2) #3}
\def\physrev#1#2#3{{Phys. Rev.} {\bf D#1} (#2) #3}
\def\ap#1#2#3{{Ann. Phys.} {\bf #1} (#2) #3}
\def\prep#1#2#3{{Phys. Rep.} {\bf #1} (#2) #3}
\def\rmp#1#2#3{{Rev. Mod. Phys. }{\bf #1} (#2) #3}
\def\rmatp#1#2#3{{Rev. Math. Phys. }{\bf #1} (#2) #3}
\def\cmp#1#2#3{{Comm. Math. Phys.} {\bf #1} (#2) #3}
\def\mpl#1#2#3{{Mod. Phys. Lett. }{\bf #1} (#2) #3}
\def\ijmp#1#2#3{{Int. J. Mod. Phys.} {\bf #1} (#2) #3}
\def\lmp#1#2#3{{Lett. Math. Phys.} {\bf #1} (#2) #3}
\def\tmatp#1#2#3{{Theor. Math. Phys.} {\bf #1} (#2) #3}
\def\hepth#1{{ hep-th/}#1}


\begin{thebibliography}{0}

   
%

 %
  \bibitem{GM}
  P. Ginsparg and G. Moore,
  ``Lectures on 2D gravity and 2D string theory (TASI 1992)", 
  hep-th/9304011.
   

 
\bibitem{DiFrancescoGinsparg}P.~Di Francesco, P.~Ginsparg and
J.~Zinn-Justin,``2-D Gravity and random matrices,'' Phys.\ Rept.\
{\bf 254} (1995) 1,  hep-th/9306153.


\bibitem{polchinski}J. Polchinski, ``What is string theory'',
{\it Lectures presented at the 1994 Les Houches Summer School
``Fluctuating Geometries in Statistical Mechanics and Field 
Theory''},  
\hepth{9411028}.

\bibitem{KlebanovMQM}I. Klebanov, {\it Lectures delivered at the ICTP
Spring School on String Theory and Quantum Gravity},
Trieste, April 1991, \hepth{9108019}.

  
   \bibitem{JevickiQN}
A.~Jevicki,
``Developments in 2-d string theory,''
 hep-th/9309115.






 %
  \bibitem{Witten}
  E.~Witten,
``Ground ring of two-dimensional string theory,''
Nucl.\ Phys.\ B { 373}, 187 (1992),
hep-th/9108004.

%
 
\bibitem{WitZw} E.~Witten and B.~Zwiebach, ``Algebraic structures
and differential geometry in 2D string theory,'' Nucl.\ Phys.\
B { 377}, 55 (1992),  hep-th/9201056.

  
  \bibitem{KlebPol}
I.~R.~Klebanov and A.~M.~Polyakov, ``Interaction of discrete
states in two-dimensional string theory,'' Mod.\ Phys.\ Lett.\ A
{6}, 3273 (1991) hep-th/9109032.



 \bibitem{KMS} D. Kutasov, E. Martinec, N. Seiberg, 
   ``Ground rings and their modules in 2-D gravity with $c\le1 $ 
matter",  
      Phys.Lett. {\bf B276} (1992) 437, \hepth{9111048}.


  
  \bibitem{bershkut}
  M. Bershadsky and D. Kutasov,
 ``Scattering of open and closed strings in (1+1)-dimensions",
 \np{382}{1992}{213}, \hepth{9204049}.
  



 \bibitem{SeibergS}
 N. Seiberg and D. Shih, ``Branes, rings and matrix models on minimal 
(super)string theory",  hep-th/0312170.
  



\bibitem{Gouli}
M. Goulian and B. Li, Phys. Rev. Lett. 66 (1991), 2051.

 
\bibitem{VDotsenko}
V. Dotsenko, Mod. Phys. Lett. A6(1991), 3601.


%
  \bibitem{berkut}
  M. Bershadsky and D. Kutasov,
  Phys. Lett. 274B (1992) 331,  hep-th/9110034.
  
 



 \bibitem{TY}Y. Tanii, S.-I. Yamaguchi, ``Two-dimensional 
   quantum gravity on a disc", Mod. Phys. Lett. A7 (1992) 521,
 hep-th/9110068; ``Disk Amplitudes in Two-Dimensional Open String 
Theories", hep-th/9203002.




 \bibitem{DO}
H. Dorn, H.J. Otto: 
Two and three point functions in Liouville theory, 
\np{429}{1994}{375}, hep-th/9403141.



\bibitem{ZZtp}A.B. Zamolodchikov, Al.B. Zamolodchikov: 
Structure constants and conformal bootstrap in Liouville field theory,
\np{477}{1996}{577}, hep-th/9506136.



 \bibitem{FZZb}V.~Fateev, A.~B.~Zamolodchikov and 
A.~B.~Zamolodchikov,
``Boundary Liouville field theory. I: Boundary state and boundary
two-point function,''  hep-th/0001012.

 
\bibitem{PTtwo}B.~Ponsot, J.~Teschner, ``Boundary Liouville Field 
Theory: Boundary three point function'',
  Nucl.~Phys.~{B622} (2002) 309, \hepth{0110244}.






 \bibitem{hosomichi}K.~Hosomichi, "Bulk-Boundary Propagator in 
Liouville 
Theory on a Disc", JHEP 0111 
 044 (2001), \hepth{0108093}.
 


%
\bibitem{newhat}
M.~R.~Douglas, I.~R.~Klebanov, D.~Kutasov, J.~Maldacena, E.~Martinec 
and N.~Seiberg,
``A new hat for the c = 1 matrix model,'' hep-th/0307195.
 

\bibitem{BGR} I. Kostov, 
 hep-th/0312301.
 


 \bibitem{KPS} I. Kostov, B. Ponsot and D. Serban,   ``Boundary  
Liouville Theory and 
2D Quantum Gravity", hep-th/0307189.
 
%




\bibitem{DF}V.S.~Dotsenko and V.A.~Fateev, "Four point correlations
functions and the operator algebra in the two dimensional
conformal invariant theories with the central charge $c<1$"
\np{251}{1985}{691}.







\bibitem{cardy}J. Cardy, ``Conformal 
invariance and surface
 critical behavior'', \np{240}{1984}{514};
``Boundary conditions, 
fusion rules and the Verlinde
 formula'', \np{324}{1989}{581}.




   
 \bibitem{kachru}S. Kachru, 
 ``Quantum Rings and Recursion Relations in 2D Quantum Gravity'',
 Mod. Phys. Lett. A7 (1992) 1419, 
  hep-th/9201072.



\bibitem{GovJcf}
S. Govindarajan, T. Jayaraman and V. John,
ÒGenus Zero Correlation Functions in
 $c<1$ String Theory,Ó Phys. Rev. D 48 (1993) 839,
\hepth{9208064}.


  \bibitem{GovLast}
  S. Govindarajan, T. Jayaraman and V. John,
  ``Correlation Functions and Multicritical Flows in $c<1$ String 
Theory",
 Int. J. Mod.Phys. A10 (1995) 477,  hep-th/9309040.
     
    

\bibitem{BPZ}
A. Belavin, A. Polyakov, A. Zamolodchikov,
  ``Infinite 
 conformal symmetry in two-dimensional quantum 
field theory", Nucl. Phys. {B241}, 333-380, (1984).


\bibitem {Teschner}
J.Teschner. On the Liouville Three-Point Function.
Phys.Lett., B363 (1995) 65. 


\bibitem{Ibliou}I. K. Kostov, ``Boundary Correlators in 2D Quantum 
Gravity:
 Liouville versus Discrete Approach ", \np{658}{2003}{397}, 
\hepth{0212194}.



 \bibitem{Idis}I. Kostov, ``Strings with discrete target space'',
 \np{376}{1992}{539},  hep-th/9112059.



 

 
\end{thebibliography}
\end{document}